\documentclass[10pt, conference, compsocconf]{IEEEtran}
\usepackage{footnote}
\usepackage{float}
\usepackage{multirow}
\usepackage{graphicx}

\usepackage[cmex10]{amsmath}
\usepackage{amsmath}
\usepackage{amssymb}
\usepackage{algorithm}
\usepackage{algorithmic}
\usepackage{xcolor}

\def\IEEEbibitemsep{0pt plus .5pt}

\begin{document}

\bstctlcite{IEEEexample:BSTcontrol}
%
\title{A Game-Theoretic Drone-as-a-Service Composition for Delivery}


\author{\IEEEauthorblockN{Babar Shahzaad\IEEEauthorrefmark{1},
Athman Bouguettaya\IEEEauthorrefmark{1},
Sajib Mistry\IEEEauthorrefmark{2}
}

\IEEEauthorblockA{\IEEEauthorrefmark{1}School of  Computer Science,
The University of Sydney, Australia\\
\{babar.shahzaad, athman.bouguettaya\}@sydney.edu.au}
\IEEEauthorblockA{\IEEEauthorrefmark{2}School of Electrical Engineering, Computing and Mathematical Sciences,
Curtin University, Australia\\
sajib.mistry@curtin.edu.au
}
}


%


\maketitle

\begin{abstract}
We propose a novel game-theoretic approach for drone service composition considering recharging constraints. We design a non-cooperative game model for drone services. We propose a non-cooperative game algorithm for the selection and composition of optimal drone services. We conduct several experiments on a real drone dataset to demonstrate the efficiency of our proposed approach.

\end{abstract}

%
\IEEEpeerreviewmaketitle

\section{Introduction}

A drone is an aircraft that can fly with no pilot onboard. Drones offer a myriad of new potential applications in several domains including agriculture, geographic mapping, healthcare, shipping, and shopping \cite{1}. Drones provide a variety of services such as \textit{inspection}, \textit{sensing}, and \textit{delivery} \cite{DBLP:journals/corr/abs-1805-00881}. Amazon and other large corporations are making efforts to commercially use drones for delivery. Drones provide \textit{faster} and \textit{cost-effective} delivery services \cite{5}. Delivery drones differ from ground vehicles which are constrained by road infrastructure and traffic congestion.

The \textit{service paradigm} \cite{Bouguettaya:2017:SCM:3069398.2983528} provides an elegant mechanism to abstract the \textit{functional} and \textit{non-functional} or \textit{Quality of Service} (QoS) properties of a drone termed as \textit{Drone-as-a-Service} (DaaS) \cite{8818436}. The \textit{functional} property represents the delivery of packages in a skyway network. The skyway network is made up of a set of line segments of which endpoints are nodes of the network. The nodes in the skyway network are the rooftops of high-rise buildings. The nodes are assumed to be \textit{delivery targets} or \textit{recharging stations}. The \textit{non-functional} properties represent the quality parameters that distinguish between functionally equivalent drone services, i.e., flight range, speed, payload, and battery capacity.

The practical utilization of drones for delivery is restricted by a number of \textit{intrinsic} and \textit{extrinsic} factors \cite{alkouz2020swarm}. The intrinsic factors include \textit{limited battery capacity}, \textit{limited flight range}, and \textit{constrained payload} of a drone. The extrinsic factors are related to the \textit{drone service environment} such as \textit{ a highly dynamic operating environment} and \textit{constraints on recharging pads} at each station. The maximum flight range of a delivery drone with full payload weight varies from 3 to 33 km \cite{12}. The \textit{battery capacity, speed, payload weight,} and \textit{weather conditions} influence the flight range of a drone \cite{10.1007/978-3-030-33702-5_28}. The drones may need multiple times of recharging to cover long-distance areas. 

The drone delivery problem is defined as the time-optimal delivery of packages from a source to a destination in a skyway network.
Using the service paradigm, we \textit{reformulate the drone delivery problem} as finding an optimal set of drone services from a given source (e.g., warehouse rooftop) to a destination (e.g., recipient’s landing pad). Therefore, the target is the selection and composition of the \textit{best} services to minimize the delivery time. DaaS composition is the process of selection of the best services that make up a skyway path from source to destination. The composition is constrained by drone payload, range, and availability of recharging stations. We assume that no handover of packages takes place among different drones at intermediate stations as this would be more realistic.

To the best of our knowledge, existing approaches mainly focus on \textit{routing} and \textit{scheduling} of drones by formulating the problem as Travelling Salesman Problem (TSP) \cite{8488559} and Vehicle Routing Problem (VRP) \cite{7513397}. Most of the studies solve the TSP and VRP problems for a combination of ground vehicles and drones \cite{Khoufi_2019}. In contrast, we focus on the DaaS composition in a \textit{dynamic multi-drone environment}, i.e., several drone services operating in the same skyway network. Each drone service has its own delivery plan. The drone services act \textit{independently} and \textit{selfishly} as they are only interested in minimizing their own delivery time. Therefore, the drone services may impact each other's composition
as they share the same network and compete for limited recharging pads which leads to congestion.

\textit{We propose a game-theoretic approach for the selection and composition of optimal DaaS services}. An optimal DaaS service avoids congestion at intermediate stations.
The composition will take into account two main constraints: (1) the availability of recharging pads at the recharging stations and (2) the influence of a drone's choice of recharging on other drones. It is assumed that the intrinsic features of each drone service are \textit{deterministic}, i.e., the payload, speed, flight range, and battery capacity of each drone are known a priori. In contrast, the extrinsic features such as the service environment are \textit{stochastic}, i.e., the availability of recharging pads is not guaranteed.



    

We predict the availability of recharging pads by considering all drone services approaching and leaving certain recharging stations during a specific time interval.
We repeat the process for all stations within the range of a selected drone service.
This process continues until the delivery of the package to the destination. We summarize the main contributions of this paper as follows:
\begin{itemize}
\item A non-cooperative game model for drone services. 

 
\item A new non-cooperative game algorithm for the selection and composition of drone services.
\item An evaluation using a real-world dataset to demonstrate the effectiveness of the proposed approach.
\end{itemize}

\section{Related Work}

Several studies address the routing and scheduling problems for drone delivery services. Most of the existing research work focuses on using drones in conjunction with ground vehicles for last-mile delivery. A hybrid framework for a traditional delivery truck and a companion drone was first studied in \cite{MURRAY201586}. Two new approaches are proposed to address the operational challenges associated with drone-assisted parcel delivery. In one of two approaches, a truck acts as a mobile depot for a drone to make deliveries to customers along its route. The drone departs from and lands back on the truck after making a delivery. In the second approach, the drone and the truck make deliveries separately. The customers that are close enough to the warehouse are served directly from the warehouse using drones. The truck makes deliveries to the customers who are not within the maximum range of the drone.
The proposed approaches use a truck to make deliveries which is not suitable for remote areas where there is no road infrastructure. \textit{The proposed approaches do not consider the recharging constraints to serve long-distance areas.}

A spatio-temporal service model is proposed for drone services in \cite{8818436}. The drone services are selected and composed considering QoS properties. The proposed model does not take into account the battery capacities of drones and recharging constraints at stations. The proposed model is extended in \cite{10.1007/978-3-030-33702-5_28} by taking into account the recharging constraints. A deterministic lookahead heuristic-based algorithm is developed to solve the proposed problem. However, the dynamic service environment, the effects of stochastic arrival of drone services on other drones, and the competition of services for the recharging pads are not considered in the proposed approach.

Game-theoretical approaches have been applied to model and analyze competing services in various areas of service computing \cite{6903269}. A non-cooperative game model is proposed to solve QoS-aware service selection and composition problem in \cite{6009423}. The game depicts a competitive relationship between concurrent tasks for multiple users. An iterative algorithm is proposed to obtain Nash equilibrium for maximizing the utility of each user by selecting a suitable service. The computational complexity for the iteration process is very high which limits the potential of the proposed approach.



A non-cooperative game is proposed to solve the recharging assignment problem of multiple robots to recharging stations in \cite{8733079}. An allocation algorithm is proposed to achieve the pure strategy Nash equilibrium. Multirobot recharging strategies are presented based on the proposed algorithm to reduce the total cost with mobile recharging stations. The efficiency of the strategies is evaluated by using the concept of the price of anarchy. The proposed approach does not consider dynamic congestion conditions at recharging stations. \textit{To our knowledge, this paper is the first attempt to model the drone delivery problem as a non-cooperative game considering the congestion conditions at recharging stations}.

\section{Non-Cooperative Game Model for Drone Services}

\subsection{Problem Statement}

We model the drone service selection and composition problem as a non-cooperative game problem. Let $P = \{p_1, p_2,\ldots, p_n\}$ be a finite set of $n$ players, each corresponding to a drone. Each player is selfish and rational and tries to maximize its own payoff which depends on its strategy (in this case is the service selection). Each selected service leads to a recharging station or a delivery target. We assume that each player knows the location information of other players without explicit communication. Let $R = \{r_1, r_2,\ldots, r_m\}$ be a finite set of $m$ resources (in this case are the recharging stations located in the network, each having a finite set of recharging pads). We assume that the recharging stations are stationary and their positions are known to the players. Each player makes a decision of recharging station selection in a non-cooperative setting. Let $t^{p_i}_{jk}$ be the travelling time of a player $p_i$ from the recharging station $r_j$ to $r_k$ (i.e., there exists a skyway segment drone service from $r_j$ to $r_k$).
Let $rt^{p_i}_k$ and $wt^{p_i}_k$ denote the recharging and waiting times of a player $p_i$ at the recharging station $k$. As the game observes \textit{FCFS} property, therefore the waiting time $wt$ of any player may vary depending on the recharging station selection by other players. We define the payoff (i.e., total time) $T^{pi}_k$ of a player $p_i$ for selecting a recharging station $k$ as follows.
\begin{equation*}
T^{pi}_k = t^{p_i}_{jk} + rt^{p_i}_k + wt^{p_i}_k \qquad (j, k \in N,\  p_i \in P)
\end{equation*}

The objective is to find an optimal set of payoffs for a player $p_i$ from a source to a destination, defined as follows.

\begin{equation*}
T_{pi} = min \sum_{k=1}^{u} T^{pi}_k \qquad (u \in N,\  p_i \in P)
\end{equation*}

where $u$ represents the total number of recharging stations where a player competes for resources and $T_{pi}$ denotes the sum of payoffs of player $p_i$.

\subsection{Non-Cooperative Game with Complete Information}

A non-cooperative game with complete information (NCG-CI) is used to solve the scheduling problems \cite{nonCop12}. We use NCG-CI as the baseline approach to finding an optimal DaaS composition plan. In this approach, there is no uncertainty involved during the composition process, i.e., the actual information about
the arrival of drone services 
is known beforehand. The NCG-CI computes all possible compositions of drone services taking into account the actual
arrival time of 
other drone services. We select an optimal DaaS composition considering the best QoS in terms of delivery time. Finding and evaluating all possible DaaS compositions in the baseline approach is computationally expensive which reduces its performance significantly.



\subsection{Drone Service Selection Game Model}

We model the drone service selection game as follows. Each player can select any service leading to the recharging stations in its range within the skyway network. The services and recharging stations are assumed to have a one-to-one correspondence. If $M$ out of total $m$ recharging stations are in the range of the player $p_i$, then player $p_i$ has $M$ possible strategies. Each strategy has a distinct payoff. Each player selects an optimal delivery service. An optimal delivery service leads to the destination faster by minimizing the travel, waiting, and recharging times of a drone.

The proposed non-cooperative game model has some flavour of a sequential game. The players' decisions are made sequentially which may influence the payoffs. The players have information about the service selections of previous players but their arrival times are not guaranteed.
The service selection of a new player may potentially influence the previous players. As a result, the decisions made previously may be invalidated by future changes. The proposed game is not a pure sequential game because the strategy (i.e., selection decision) of one player may cause multiple players to change their strategies.

In general, the optimal service selection and composition in a resource-constrained environment is an NP-hard problem \cite{7558006}. We define some reasonable assumptions to solve the problem in polynomial time. 
We assume that a desired player $p_i$ has the following information about recharging stations in its range for decision-making at a particular station. Statically, the desired player $p_i$ is given that (1) How many drones are being recharged at each particular station, (2) How long each drone still needs to get recharged, (3) How many other drones are waiting to be recharged, and (4) How many drones are expected to reach each particular station.

\section{Solution Algorithm for Non-Cooperative Game}

\begin{figure}

    \centering

    \includegraphics[width=0.49\textwidth]{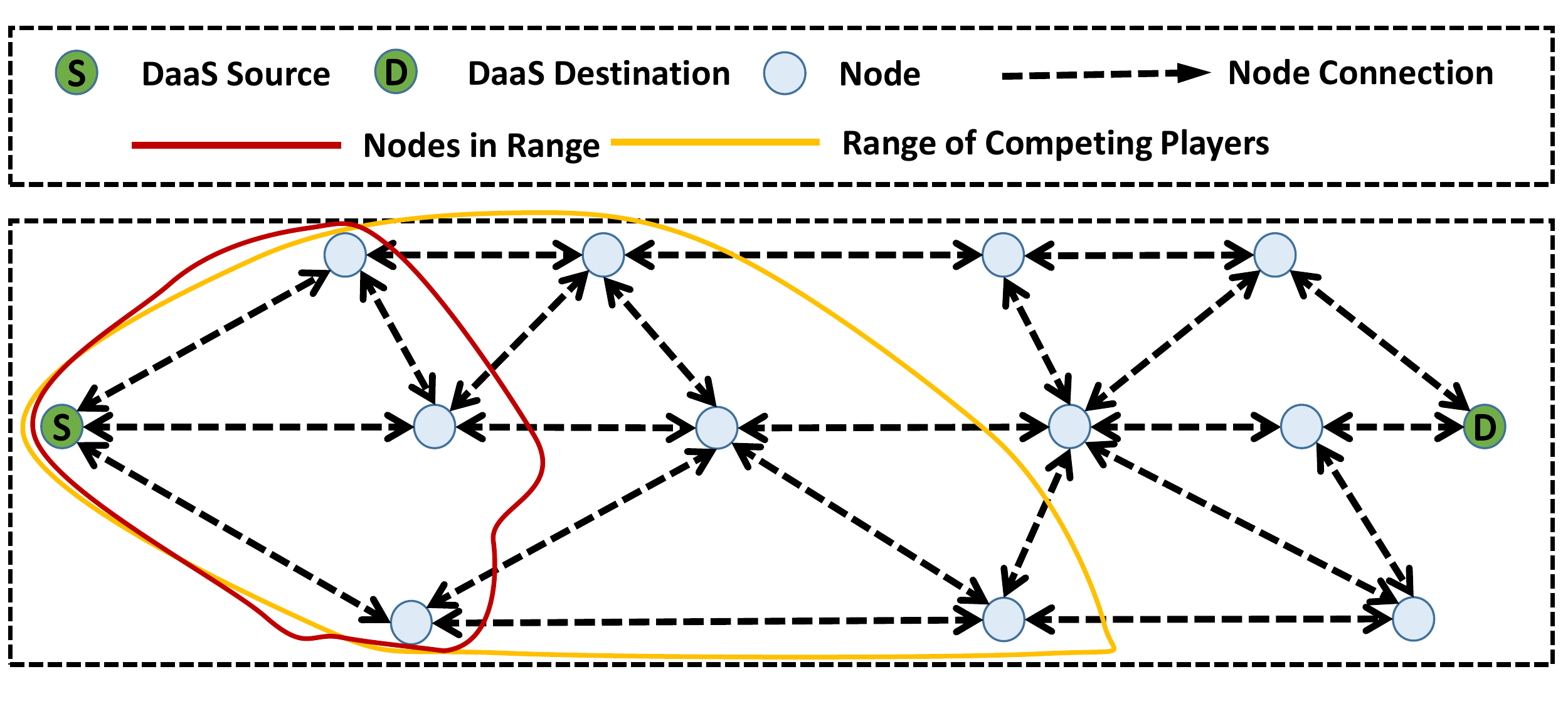}
    
    \caption{Local skyway network for reducing search space}
    \label{fig2}

\end{figure}

In a non-cooperative game, the objective of each player is simply to maximize its reward. The players present myopic behaviour to minimize the total delivery time, i.e., each player makes a greedy decision based on the current information. The greedy approach may not provide the best outcome due to the congestion at certain nodes and
the lack of real-time communication with other players.

We present a prediction-based non-cooperative game (NCG-PB) theory approach to deal with the congestion at certain nodes.
Most of the existing approaches
focus on finding the shortest path to the destination. The travel time for the shortest path may not guarantee the overall shortest time to the destination due to the congestion conditions at intermediate recharging stations. We, therefore, consider the summation of the shortest travel time, waiting time, and recharging time to select a particular service.

The search space becomes huge if we compute strategies for all players in the skyway network. The desired player can reach a limited number of recharging stations. We reduce the search space by only considering players that can arrive at stations in the range of the desired player, termed as local skyway network. For example, a desired player $p_i$ can reach a recharging station $r_j$ at 10:00 am. We consider only those players that are expected to reach the station $r_j$ between 09:50 am to 10:10 am. We assume this threshold to incorporate the variations in arrival times of other players. Fig. \ref{fig2} illustrates the local skyway network for reducing search space.
\begin{algorithm}[t]
\small
\caption{Non-Cooperative Game Algorithm}\label{alg:algorithm1}
	\begin{algorithmic}[1]
    \REQUIRE
		$G$, $src$, $dst$, $R$, $P$, $w$, $tw$
	\ENSURE
		$Comp$\\
		\STATE $Comp \gets \phi$
		\STATE $Time \gets \phi$
		\STATE $St_{range} \gets \phi$
		\STATE $p_i \gets P[i]$
		\STATE $curLoc \gets src$
		\STATE $St_{range} \gets$ get stations in range of player $p_i$ from $curLoc$
		\WHILE{$St_{range}$ is not empty}
		\IF{$dst$ is in $St_{range}$}
		\STATE compute $Time$ to $dst$ from $curLoc$
		\STATE $DaaS \gets$ a segment from $curLoc$ to $dst$
		\STATE $Comp$.append ($DaaS$)
		\RETURN $Comp$
		\ELSE
		\STATE get list of scheduled players from $P$ at each $st$ in $St_{range}$ in time $tw$
		\STATE compute travel, wait, and recharge times to each $st$ from $curLoc$
		\STATE compute distance to $dst$ from each $st$
		\STATE select $st$ with minimum $Time$ from $curLoc$ and time to $dst$ 
		\STATE $DaaS \gets$ a segment from $curLoc$ to $st$
		\STATE $Comp$.append($DaaS$)
		\STATE $curLoc \gets st$ 
		\STATE select $St_{range}$ from $curLoc$
		\ENDIF
		\ENDWHILE
	\end{algorithmic}
\end{algorithm}
\setlength{\textfloatsep}{15pt}
\subsection{Algorithm}

We present a new non-cooperative game algorithm for drone service selection. The service selection at each recharging station leads to the composition of drone services from source to destination. The details of the proposed algorithm are presented in Algorithm \ref{alg:algorithm1}. In Algorithm \ref{alg:algorithm1}, the output is a composition delivery plan from a given source to a destination. It takes as input a skyway network given by Graph $G$, the source $src$, the destination $dst$, a set of resources (recharging stations) $R$, a set of players $P$,
the weight of package $w$,
the time window $tw$ for choosing the competing players.
We create empty lists for the output composition plan $Comp$, the total delivery time $Time$, and the recharging stations in the range of the desired player $St_{range}$ (Lines 1-3). We choose a player $p_i$ from a given set of players $P$ in the skyway network (Line 4). We assume that the following attributes for each player in $P$ are given: the speed, the flight range, and the recharging time. We get all stations in the range of player $p_i$ depending on the package weight $w$ (Line 6). If the player $p_i$ can reach the destination without requiring recharging, then there is no competition with other players. It directly moves from source to destination, returning a segment from the given source to destination (Lines 8-12). If the player $p_i$ requires recharging to reach the destination, we compute the competing players and their scheduled arrival times $schedPls$ at each station in the range of player $p_i$. We consider only players arriving at any particular station of player $p_i$ within a certain time window $tw$ (Line 14).
The travel time, waiting time, and recharging time are computed for each station in the range of player $p_i$ (Line 15). We compute the distances from all the nearby stations to the given destination $dst$ of player $p_i$ (Line 16).
We select an adjacent station from the current location of player $p_i$ which takes the overall shortest travel time (Line 17). On each iteration, we simply add the selected skyway segment service to the composition plan $Comp$ (Lines 18-19). We update the current location and then search for the nearby recharging stations of the player $p_i$ which in turn updates its neighbour services (Lines 20-21). This process continues till the destination node is discovered or nearby stations list is empty (i.e., no service composition plan found).

\section{Experiments and Results}
We perform several experiments to investigate the performance of the proposed game-theoretic approach. We compare the proposed prediction-based non-cooperative game (NCG-PB) theory approach with a non-cooperative game with complete information (NCG-CI) \cite{nonCop12} approach.
We focus on the run-time complexity and average delivery time for evaluating the proposed solution approach.

\subsection{Experimental Settings}
In this section, we explain the setup of the simulation environment. We use NetworkX python library to build the topology of the skyway network. We model the delivery drones operating in the skyway network. We evaluate the proposed NCG-PB approach using a real drone dataset \cite{14}. The dataset contains the trajectories of drones, which include data for coordinates, altitude, and timestamps. 
Table \ref{tab:table1} summarizes the simulation parameters and values.
We conduct experiments for an average of 50 percent times the total nodes. For example, an experiment is performed 50 times for 100 nodes. Each experiment starts with a random source and a destination point.

\begin{table}[t]
\centering
\scriptsize
\caption{Experiment Variables}
\label{tab:table1}
\begin{tabular}{|p{5cm}|l|}
\hline
 Simulation Parameter &  Value \\

\hline


Drone name &  DJI M200 V2\\ \hline

Payload & 1.45 Kg \\ \hline

Flight time & 24 min \\ \hline

Flight range & 32.4 km \\ \hline

Max speed & 81 km/h \\ \hline

Recharging time & 60 min \\ \hline

Battery capacity & 4280 mAh \\ \hline

Number of nodes &  300  \\ \hline

Experiment run (\% times the total nodes) & 50 \\

\hline
\end{tabular}
\end{table}


\subsection{Results and Discussion}

The proposed game-theoretic approach performs the composition of selective drone services considering the congestion conditions at intermediate recharging stations.
The target is to reach the destination faster.




    




    


\subsubsection{Run-time Complexity}

\begin{figure}[t]
    \centering
        \centering
        \includegraphics[width=0.4\textwidth]{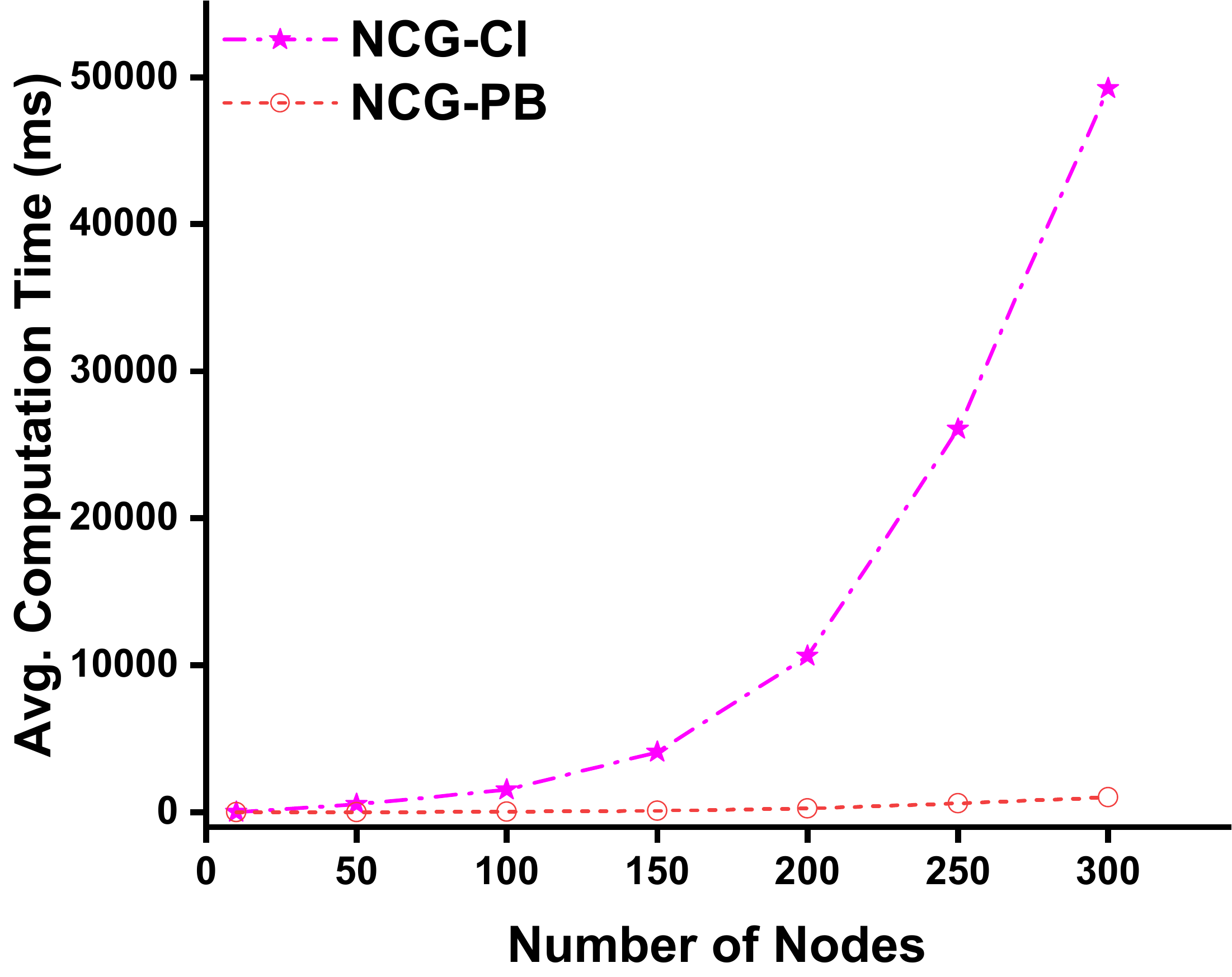} 
        \caption{Average computation time}
        \label{fig:fig4.pdf}
    \vspace{-0.3cm}
\end{figure}

The run-time complexity of the NCG-CI approach is very high compared to
our proposed NCG-PB approach. The computation time increases due to the increasing number of possible compositions for drone services. The difference in run-time complexity for NCG-CI and NCG-PB approaches is shown in  Fig.~\ref{fig:fig4.pdf}. As expected, the average computation time for the increasing number of nodes is much higher for the NCG-CI approach than our proposed approach. It is impractical to use the baseline approach in real-world scenarios as it is exhausted for large scale problems.

\subsubsection{Average Delivery Time}

\begin{figure}[t]
    \centering
        \centering
        \includegraphics[width=0.4\textwidth]{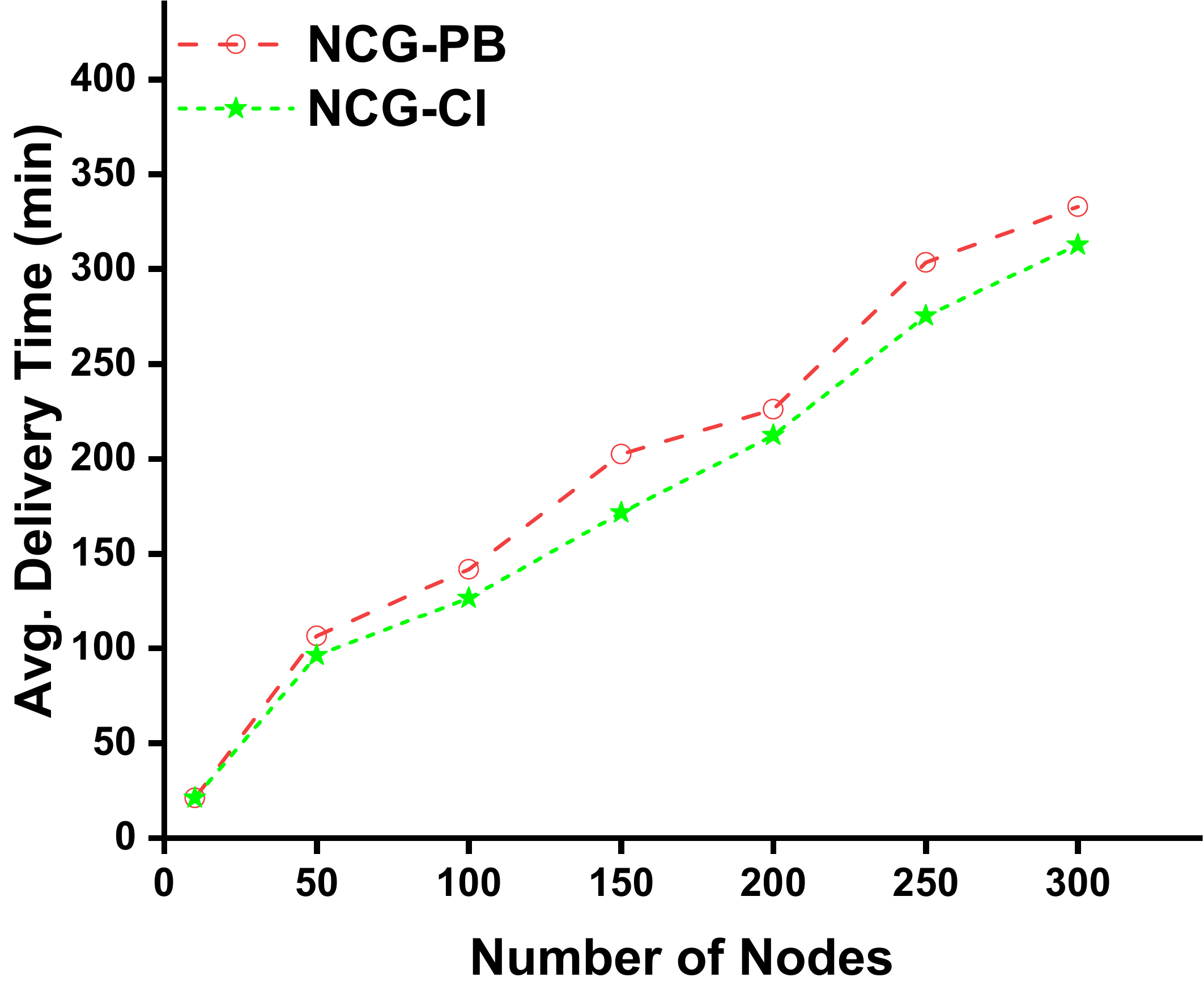} 
        \caption{Average delivery time}
        \label{fig:fig5.pdf}
    \vspace{-0.3cm}
\end{figure}

The delivery time is highly uncertain for composite drone services. The competing drone services can occupy recharging pads for long time periods. The delivery time includes the flight time, waiting time, and recharging time at each station. The selection of a \emph{right} drone service guarantees the availability of recharging pads ahead of time which minimizes the overall delivery time.
Fig.~\ref{fig:fig5.pdf} shows the efficiency of the proposed NCG-PB approach compared to NCG-CI baseline approach. The NCG-CI approach provides the exact solution as it finds all possible compositions.
Our proposed NCG-PB approach provides computationally fast and near-optimal solutions for delivering the packages compared to
the NCG-CI approach.


\section{Conclusion}

We propose a game-theoretic approach for drone service composition considering recharging constraints. The proposed NCG-PB algorithm significantly reduces the search space of candidate drone service compositions. It enables a fast online composition plan which is essential for time-constrained drone-based delivery requirements. Experimental results illustrate that the proposed approach produces near-optimal DaaS compositions compared to the NCG-CI baseline approach. In future, we plan to consider handovers at intermediate stations and uncertain wind effects on the battery consumption of drones.

\section*{Acknowledgment}
This research was partly made possible by DP160103595 and LE180100158 grants from the Australian Research Council. The statements made herein are solely the responsibility of the authors.

\bibliographystyle{IEEEtran}
\bibliography{references}

\end{document}